\documentclass[aps,prd,showpacs,twocolumn
,superscriptaddress]{revtex4}
\usepackage{graphicx}% Include figure files
\usepackage{dcolumn}% Align table columns on decimal point
\usepackage{bm}% bold math
\usepackage{float}
\usepackage{color}
\usepackage[normalem]{ulem} % \sout{old text} for strikeout
\usepackage[dvipsnames]{xcolor} % For blue in-text comments
\usepackage{hyperref}
\hypersetup{
  colorlinks=true,        % false: boxed links; true: colored links
  linkcolor=blue,         % color of internal linksx
  citecolor=cyan,         % color of links to bibliography
}

\usepackage{doi}
\usepackage{amsmath}
\usepackage{orcidlink}
\usepackage{amssymb}

\usepackage{booktabs}

\begin{document}

\title{Quasibound states of charged Dirac fields around\\ Einstein--Euler--Heisenberg black holes}

\author{Nodirbek Anarboev\orcidlink{0009-0009-6167-5216}}
\email{nodirbekanarboyev@gmail.com}
\affiliation{Institute of Theoretical Physics, National University of Uzbekistan, Tashkent 100174, Uzbekistan}

\author{Javlonbek Khasanov\orcidlink{0009-0000-2772-3960}}
\email{khasanovjavlonbek20@gmail.com}
\affiliation{Institute of Theoretical Physics, National University of Uzbekistan, Tashkent 100174, Uzbekistan}

\author{Javlon Rayimbaev\orcidlink{0000-0001-9293-1838}}
\email{javlonrayimbaev6@gmail.com}
\affiliation{Institute of Theoretical Physics, National University of Uzbekistan, Tashkent 100174, Uzbekistan}
\affiliation{Tashkent State Technical University, Tashkent 100095, Uzbekistan}

\author{Bobomurat Ahmedov\orcidlink{0000-0002-1232-610X}} 
\email{ahmedov@astrin.uz}
\affiliation{Institute of Theoretical Physics, National University of Uzbekistan, Tashkent 100174, Uzbekistan}
\affiliation{School of Physics, Harbin Institute of Technology, Harbin 150001, People’s Republic of China}
\affiliation{Institute for Advanced Studies, New Uzbekistan University, Movarounnahr str. 1, Tashkent 100000, Uzbekistan}

\begin{abstract}
We study quasibound states of a massive charged Dirac field on the Einstein--Euler--Heisenberg (EEH) black hole, the extension of Reissner--Nordstr\"om (RN) by the leading weak-field vacuum-polarization term of QED. We derive the background to first order in the Euler--Heisenberg coupling $a$. We show that it admits a nondegenerate horizon at $Q=M$ and an extremal charge $Q_{\rm ext}=M[1+a/(40M^2)+\mathcal{O}(a^2)]>M$, with horizon data that depend on $\sqrt a$ near extremality. From the separated radial equations, we derive an identity that excludes real-frequency quasibound states on any background of this class, so damped branches cannot become unstable. Since the EEH corrections first enter at $\mathcal{O}(r^{-5})$ in the electrostatic potential, the trapping condition $M\mu^2-qQ\omega_R>0$ and the hydrogenic spectrum coincide with RN, and no fine-structure-order splitting arises. Two-sided shooting with a Muller root finder shows that the real frequencies shift only by near-zone amounts that are linear in $a$ and strongly suppressed with angular momentum. The damping rates respond in a branch-dependent way. At $Q/M=0.99$ the $\ell=0$ ground state becomes slightly longer-lived, while the $(n_r,j,\ell,\lambda)=(0,\tfrac32,1,-2)$ state decays up to $80\%$ faster for $a/M^2\le1$, and near the trapping threshold the damping rate increases by up to $60\%$. Switching on the metric and potential corrections separately, together with a WKB analysis of the inner barrier, identifies the mechanisms: Coulomb screening always increases leakage, while the metric correction controls the sign and size of the $\ell$-dependent effect. Characteristic time-domain evolutions with Richardson-extrapolated matrix-pencil extraction reproduce the frequency-domain shifts.
\end{abstract}
%\pacs{04.50.-h, 04.40.Dg, 97.60.Gb}

\maketitle

%========================================================================
\section{Introduction}
\label{sec:intro}
%========================================================================

Massive fields whose Compton wavelength is comparable to the gravitational radius can be trapped outside a black hole by the mass barrier. The resulting quasibound states slowly leak through the horizon~\cite{Detweiler1980,Dolan2007,KonoplyaZhidenkoRMP,BritoCardosoPani}. Their complex frequencies $\omega=\omega_R+i\omega_I$ ($\omega_I<0$) carry two different kinds of information. The real part is fixed mostly by the long-range Newtonian and Coulomb tails, while the damping rate measures how efficiently the cloud reaches the horizon. Fermionic quasibound states have been studied in Schwarzschild, Reissner--Nordstr\"om (RN) and Kerr spacetimes~\cite{Lasenby2005,DolanDempsey2015,Huang2017,Sporea2019,KonoplyaZhidenko2018}. For charged massive Dirac fields on RN they form long-lived hydrogen-like spectra~\cite{Huang2017,Jing2005}. Unlike bosons, Dirac fields are never superradiantly amplified~\cite{Unruh1976,IyerKumar1978,Finster2000}, so their damping rates cleanly probe the near-horizon structure.

Recently, Ref.~\cite{Zhang2026} studied charged Dirac quasibound states on the regular Ay\'on-Beato--Garc\'ia (ABG) black hole~\cite{AyonBeatoGarcia1998}, a solution of Einstein gravity coupled to a nonlinear electrodynamics (NLED) designed to remove the central singularity. Two features were found. The real frequencies differ from RN only at fine-structure order, because the ABG potential deviates from Coulomb at $\mathcal{O}(r^{-2})$. The damping rates can differ by more than an order of magnitude because the inner potential barrier changes.

In this work, we ask the same question for a physically better-motivated NLED: the Euler--Heisenberg (EH) effective action of QED~\cite{HeisenbergEuler1936,Schwinger1951,Dunne2004}, which encodes the leading vacuum-polarization correction to Maxwell theory. Coupled to gravity, it yields the Einstein--Euler--Heisenberg (EEH) black hole~\cite{YajimaTamaki2001,MagosBreton2020,BretonLammerzahlMacias2019}. This background differs from the ABG case in three ways that are important for fermionic clouds.
\begin{enumerate}
\item The EEH corrections to the metric and to the electrostatic potential begin at $\mathcal{O}(r^{-6})$ and $\mathcal{O}(r^{-5})$, respectively. The far zone is therefore RN to a much higher order than for ABG, and no far-zone fine-structure splitting appears.
\item The EH coupling is not a regularization device but a one-parameter deformation of RN, with RN recovered continuously at $a=0$. All differences can therefore be followed as functions of a single coupling.
\item Vacuum polarization modifies the extremal limit. The EEH black hole admits a nondegenerate horizon at $Q=M$ and exists up to $Q_{\rm ext}>M$, with a nonanalytic $\sqrt a$ dependence of the horizon radius and surface gravity near extremality.
\end{enumerate}

Our main results are the following. (i) We derive the EEH solution in a self-contained way to first order in the EH coupling, obtain the extremal charge $Q_{\rm ext}/M=1+a/(40M^2)+\dots$ and the near-extremal horizon data in closed form, and quantify the weak-field validity of the parameter range used. (ii) From the first-order Dirac radial system, we derive a current identity which shows, for any static background of this type, that real-frequency quasibound states do not exist, so damped branches can never cross into instability. (iii) Using two-sided shooting with a Muller root finder, we compute the spectrum and compare EEH with RN at identical parameters. The real-frequency shifts are tiny, linear in $a$, and suppressed roughly as $(M\mu\,\alpha_{\rm eff})^{2\ell+3}$, while the damping-rate changes depend strongly on the spinor branch and are largest for the higher-$j$ mode. At $Q/M=0.99$ and $a/M^2=1$, the damping rate of the $\ell=0$ ground state changes by less than $3\%$, but that of the $\ell=1$, $j=3/2$ mode changes by $80\%$. (iv) Artificially switching off either the metric or the potential correction shows which piece of the EH physics controls each branch, and a WKB estimate of the inner barrier accounts for the $\ell=1$ trends. (v) Characteristic time-domain evolutions with matrix-pencil mode extraction independently confirm the frequency-domain results. For this check, we use Richardson extrapolation to remove the Crank--Nicolson scheme's second-order phase error.

The paper is organized as follows. Section~\ref{sec:background} derives the EEH background and its horizon structure. Section~\ref{sec:dirac} sets up the Dirac problem, the boundary conditions, the current identity, and the far-zone estimate. Section~\ref{sec:method} describes the numerical method and its checks. Section~\ref{sec:results} presents the frequency-domain results, and Sec.~\ref{sec:timedomain} the time-domain analysis. Section~\ref{sec:conclusions} summarizes the results. We use $G=c=\hbar=4\pi\epsilon_0=1$, the Fourier convention $e^{-i\omega t}$ and the gauge $A_t(\infty)=0$, and we set $M=1$ in numerical work.

%========================================================================
\section{Einstein--Euler--Heisenberg black holes}
\label{sec:background}
%========================================================================

\subsection{Field equations and the static solution}

We consider Einstein gravity minimally coupled to a nonlinear electrodynamics (NLED) Lagrangian $\mathcal{L}(\mathcal{F},\mathcal{G})$,
\begin{equation}
  S=\frac{1}{16\pi}\int d^4x\,\sqrt{-g}\,\Big[R+4\mathcal{L}(\mathcal{F},\mathcal{G})\Big],
  \label{eq:action}
\end{equation}
$\mathcal{F}=\tfrac14F_{\mu\nu}F^{\mu\nu}$, $\mathcal{G}=\tfrac14\tilde F_{\mu\nu}\tilde F^{\mu\nu}$ in units $G=c=\hbar=4\pi\epsilon_0=1$. The Maxwell theory corresponds to $\mathcal{L}=-\mathcal{F}$, for which \eqref{eq:action} reduces to the standard Einstein--Maxwell action in Gaussian units. The weak-field Euler--Heisenberg (EH) effective Lagrangian of QED~\cite{HeisenbergEuler1936,Schwinger1951,Dunne2004} is
\begin{equation}
  \mathcal{L}_{\rm EH}=-\mathcal{F}+\frac{a}{2}\mathcal{F}^2+\frac{7a}{8}\mathcal{G}^2 ,
  \label{eq:LEH}
\end{equation}
where the coupling $a>0$ has dimensions of length squared. For the purely electric, static, and spherically symmetric configurations considered here, $\mathcal{G}=0$, so only the $\mathcal{F}^2$ term contributes~\cite{YajimaTamaki2001,MagosBreton2020}.

We take the metric ansatz
\begin{equation}
  ds^2=-f(r)\,dt^2+\frac{dr^2}{f(r)}+r^2\big(d\theta^2+\sin^2\theta\,d\phi^2\big),
  \label{eq:metric-ansatz}
\end{equation}
$f(r)=1-\cfrac{2m(r)}{r}$ and an electrostatic potential $A_\mu=(A_t(r),0,0,0)$. Defining the radial field $E(r)\equiv A_t'(r)$, the only nonvanishing invariant is
\begin{equation}
  \mathcal{F}=\tfrac12 F_{tr}F_{tr}\,g^{tt}g^{rr}=-\tfrac12E^2 .
\end{equation}

\paragraph{Gauss law.}
The NLED field equations $\nabla_\mu\!\left(\mathcal{L}_{\mathcal F}F^{\mu\nu}\right)=0$, with $\mathcal{L}_{\mathcal F}\equiv\partial\mathcal{L}/\partial\mathcal{F}=-1+a\mathcal{F}=-1-\tfrac{a}{2}E^2$, reduce for the ansatz above to $\partial_r\!\left(r^2\mathcal{L}_{\mathcal F}E\right)=0$. Fixing the integration constant by the asymptotic charge $Q$ gives the cubic constraint
\begin{equation}
  r^2E\left(1+\frac{a}{2}E^2\right)=Q .
  \label{eq:gauss}
\end{equation}
The EH Lagrangian \eqref{eq:LEH} is itself the first term of a weak-field expansion, so we solve \eqref{eq:gauss} consistently to first order in $a$:
\begin{equation}
  E(r)=\frac{Q}{r^2}-\frac{a}{2}\frac{Q^3}{r^6}+\mathcal{O}(a^2).
  \label{eq:E}
\end{equation}
The field is reduced relative to the Coulomb value, as expected for a vacuum that behaves as a nonlinear dielectric with permittivity $1+aE^2/2>1$.

\paragraph{Energy density and mass function.}
For $\mathcal{L}(\mathcal{F})$ the energy--momentum tensor that follows from \eqref{eq:action} is
\begin{equation}
  T_{\mu\nu}=\frac{1}{4\pi}\Big(\mathcal{L}\,g_{\mu\nu}-\mathcal{L}_{\mathcal F}F_{\mu\alpha}F_{\nu}{}^{\alpha}\Big),
\end{equation}
which reduces to the Maxwell tensor for $\mathcal{L}=-\mathcal{F}$. Using $F_{t\alpha}F^{t\alpha}=-E^2$, the energy density is
\begin{equation}
  \rho=-T^t{}_t=\frac{1}{4\pi}\Big(-\mathcal{L}-E^2\mathcal{L}_{\mathcal F}\Big)
     =\frac{1}{4\pi}\left(\frac{E^2}{2}+\frac{3a}{8}E^4\right).
\end{equation}
Inserting \eqref{eq:E} and keeping $\mathcal{O}(a)$,
\begin{equation}
  4\pi\rho=\frac{Q^2}{2r^4}-\frac{aQ^4}{8r^8}+\mathcal{O}(a^2).
\end{equation}
The $tt$ component of the Einstein equations is $m'(r)=4\pi r^2\rho$, which integrates to
\begin{equation}
  m(r)=M-\frac{Q^2}{2r}+\frac{aQ^4}{40\,r^5},
\end{equation}
where $M$ is the ADM mass. The metric function and the potential (in the gauge $A_t(\infty)=0$) are therefore
\begin{align}
  f(r)&=1-\frac{2M}{r}+\frac{Q^2}{r^2}-\frac{aQ^4}{20\,r^6},
  \label{eq:fEEH}\\
  A_t(r)&=-\frac{Q}{r}+\frac{aQ^3}{10\,r^5}.
  \label{eq:AtEEH}
\end{align}
We have verified symbolically that the angular Einstein equation $G^\theta{}_\theta=8\pi T^\theta{}_\theta=2\mathcal{L}$ is satisfied by \eqref{eq:fEEH} and \eqref{eq:E} to the same order. Equation \eqref{eq:fEEH} is the Einstein--Euler--Heisenberg (EEH) black hole of Refs.~\cite{YajimaTamaki2001,MagosBreton2020}; for $a=0$ it reduces to Reissner--Nordstr\"om (RN). Since the RN geometry is the $a\to0$ member of the same family, we compare every EEH result below with RN at identical $(M, Q,\mu,q)$ and angular labels, so all differences are attributable to EH vacuum polarization alone.

At large radius, the EH terms are of relative order $r^{-4}$:
\begin{equation}
  f(r)=1-\frac{2M}{r}+\frac{Q^2}{r^2}+\mathcal{O}(r^{-6}),\quad
  A_t(r)=-\frac{Q}{r}+\mathcal{O}(r^{-5}).
  \label{eq:asympt}
\end{equation}
In contrast with regular black holes of the Ay\'on-Beato--Garc\'ia (ABG) type, whose potential already deviates from RN at order $r^{-2}$~\cite{AyonBeatoGarcia1998,Zhang2026}, the EEH far zone is RN to a very high order. This property organizes all the spectral results of this paper.

\subsection{Horizon structure and the extremal limit}
\label{sec:horizons}

Horizons are the positive roots of
\begin{equation}
  P(r)\equiv r^6f(r)=r^6-2Mr^5+Q^2r^4-\frac{aQ^4}{20}.
  \label{eq:P}
\end{equation}
Because $P(0)=-aQ^4/20<0$ for $a>0$, there is always at least one horizon, and $f\to-\infty$ as $r\to0^+$. The event horizon $r_h$ is the largest root, and its surface gravity is $\kappa=f'(r_h)/2$; below we use $f'(r_h)$, which enters the near-horizon exponents directly.

\paragraph{Degenerate horizon.} The outer horizon becomes degenerate when $P=P'=0$. The second condition, $P'(r)=2r^3\left(3r^2-5Mr+2Q^2\right)=0$, gives
\begin{equation}
  r_{\rm ext}=\frac{5M+\sqrt{25M^2-24Q^2}}{6},
  \label{eq:rext}
\end{equation}
and the first condition then fixes the extremal coupling
\begin{equation}
  a_{\rm ext}(Q)=\frac{20\,r_{\rm ext}^4\left(r_{\rm ext}^2-2Mr_{\rm ext}+Q^2\right)}{Q^4}.
  \label{eq:aext}
\end{equation}
Expanding \eqref{eq:aext} around $Q=M$ with $Q=M(1+\delta)$ gives $a_{\rm ext}/M^2=40\delta-460\delta^2+\mathcal{O}(\delta^3)$. Inverting this series, the maximal charge that still admits a nondegenerate outer horizon is
\begin{equation}
\begin{aligned}
  &\frac{Q_{\rm ext}}{M}=1+\frac{a}{40M^2}+\frac{23}{3200}\left(\frac{a}{M^2}\right)^2+\mathcal{O}(a^3),\\
  &\frac{r_{\rm ext}}{M}=1-\frac{a}{10M^2}+\mathcal{O}(a^2).
\end{aligned}
  \label{eq:Qext}
\end{equation}
Vacuum polarization therefore \emph{raises} the extremal charge above the RN bound $Q=M$. The exact values obtained from \eqref{eq:rext}--\eqref{eq:aext} are $Q_{\rm ext}/M=1.002576$, $1.006780$ and $1.015220$ for $a/M^2=0.1$, $0.25$ and $0.5$, in agreement with \eqref{eq:Qext} to $\mathcal{O}(a^3)$. For $Q>Q_{\rm ext}$, the outer branch disappears, and the only horizon lies deep inside, where the weak-field expansion no longer holds (see below); we do not consider that region.

\paragraph{The point $Q=M$.} At $Q=M$, where RN is extremal and $f'(r_h)=0$, the EEH horizon is nondegenerate. Writing $r_h=M(1+s)$ and solving $P=0$ order by order in $\sqrt a$,
\begin{equation}
\begin{aligned}
  \frac{r_h}{M}\Big|_{Q=M}=1+\sqrt{\frac{a}{20M^2}}-\frac{a}{10M^2}+\mathcal{O}(a^{3/2}),\\
  M f'(r_h)\big|_{Q=M}=\sqrt{\frac{a}{5M^2}}-\frac{a}{5M^2}+\mathcal{O}(a^{3/2}).
\end{aligned}
  \label{eq:QeqM}
\end{equation}
The nonanalytic $\sqrt a$ dependence shows that the near-extremal region is where the EH coupling has its strongest effect on the horizon geometry: a perturbation of order $a$ in the metric produces an $\mathcal{O}(\sqrt a)$ change in $r_h$ and in the surface gravity. For $a/M^2=0.1$, \eqref{eq:QeqM} gives $r_h/M=1.0607$ and $Mf'(r_h)=0.1214$; the exact roots of \eqref{eq:P} give $1.0626$ and $0.1240$.

\subsection{Range of validity and physical scale of $a$}
\label{sec:validity}

The truncation \eqref{eq:LEH} and the first-order solution \eqref{eq:E}--\eqref{eq:AtEEH} require $aE^2/2\ll1$ in the region probed by the field. We monitor the horizon value
\begin{equation}
  \varepsilon_h\equiv\frac{a}{2}E(r_h)^2\simeq\frac{aQ^2}{2r_h^4},
  \label{eq:epsh}
\end{equation}
which is the largest value outside the black hole. In the parameter range used below ($Q/M\le0.99$, $a/M^2\le1$) we have $\varepsilon_h\le0.23$; at the reference point $Q/M=0.99$, $a/M^2=0.5$ one finds $\varepsilon_h=0.127$. The extremal configurations of \eqref{eq:Qext} have $\varepsilon_h=0.05$, $0.14$, and $0.36$ for $ a/M^2=0.1$, $0.25$, and $0.5$, so the largest coupling should be regarded as qualitative near $Q_{\rm ext}$.

With the normalization \eqref{eq:action}, the QED coupling is $a=4\alpha/(45\pi E_c^2)$, where $E_c=m_e^2c^3/(e\hbar)$ is the Schwinger critical field in geometrized units. This gives $a/M^2\simeq5.9\times10^{7}\,(M_\odot/M)^2$, so $a/M^2=\mathcal{O}(1)$ corresponds to $M\sim8\times10^3M_\odot$. However, \eqref{eq:epsh} can be written $\varepsilon_h=(2\alpha/45\pi)(E/E_c)^2$, so $\varepsilon_h\sim0.1$ already requires $E\simeq30E_c$. Such near-extremal charges are far above the Schwinger threshold and would be rapidly neutralized by pair production. As in most of the EEH literature~\cite{YajimaTamaki2001,MagosBreton2020,BretonLammerzahlMacias2019}, we therefore treat $a$ as an effective NLED coupling and $Q$ as a free parameter. Read the results as a controlled study of how a weak-field vacuum-polarization correction to the black-hole exterior shapes the fermionic spectrum, not as a prediction for astrophysical charged black holes.

%========================================================================
\section{Charged Dirac field on the EEH background}
\label{sec:dirac}
%========================================================================

\subsection{Separated radial system}

A massive charged Dirac field $\Psi$ with mass $\mu$ and charge $q$ obeys $(\gamma^\mu D_\mu-\mu)\Psi=0$ with $D_\mu=\partial_\mu-\Gamma_\mu-iqA_\mu$. Separation in a general static metric of the form \eqref{eq:metric-ansatz} with an electrostatic potential is standard~\cite{Chandrasekhar1983,DolanDempsey2015,Huang2017,Zhang2026}. With the decomposition $\Psi\propto e^{-i\omega t}e^{im\phi}$ and the diagonal tetrad, the angular problem is that of spin-weighted spinor harmonics with separation constant
\begin{equation}
  \lambda_{j\ell}=
  \begin{cases}
    +(j+\tfrac12), & \ell=j+\tfrac12,\\[2pt]
    -(j+\tfrac12), & \ell=j-\tfrac12,
  \end{cases}
  \qquad j=\tfrac12,\tfrac32,\dots
  \label{eq:lambda}
\end{equation}
The radial functions $R_{1,2}(r)$ satisfy the first-order system
\begin{equation}
\begin{aligned}
  &\frac{dR_1}{dr}=\frac{iK}{\Delta}R_1+\frac{\lambda+i\mu r}{\sqrt\Delta}R_2,\\
  &\frac{dR_2}{dr}=-\frac{iK}{\Delta}R_2+\frac{\lambda-i\mu r}{\sqrt\Delta}R_1,
\end{aligned}
  \label{eq:radial}
\end{equation}
with
\begin{equation}
  \Delta(r)=r^2f(r),\qquad K(r)=r^2\big[\omega+qA_t(r)\big].
\end{equation}
The background enters \eqref{eq:radial} only through $f$ and $A_t$, so the system is formally identical to the RN and ABG cases~\cite{Huang2017,Zhang2026} after the replacements \eqref{eq:fEEH}--\eqref{eq:AtEEH}. Introducing the tortoise coordinate $dx/dr=1/f$ and $\psi_1=R_1/F$ with $F=\Delta^{1/4}r^{-1}(1+i\mu r/\lambda)^{1/2}$ gives the Schr\"odinger-like form $d^2\psi_1/dx^2=V_1\psi_1$, with
\begin{align}
  V_1&=-\frac{\Delta}{r^4}U_1-\frac{\Delta'^2}{8r^4}+\frac{\Delta\Delta'}{2r^5}-\frac{\Delta^2F''}{r^4F}+\frac{\Delta^2}{2r^5}\frac{\mu(\mu r-2i\lambda)}{(\lambda+i\mu r)^2},
  \label{eq:V1}\\
  U_1&=\frac{1}{\Delta}\Big(K^2+\tfrac{i}{2}K\Delta'\Big)-iK'-\frac{\mu K}{\lambda+i\mu r}-(\lambda^2+\mu^2r^2),
\end{align}
where primes denote $d/dr$. In the convention $\psi''=V\psi$, regions with $\operatorname{Re}V_1>0$ act as barriers.

\subsection{Boundary conditions and the trapping condition}

Near the horizon, \eqref{eq:radial} has a regular singular point. The ingoing Frobenius branch is
\begin{equation}
\begin{aligned}
  &R_2\simeq(r-r_h)^{\rho_h},\quad R_1\simeq a_0(r-r_h)^{\rho_h+1/2},\\
  &\rho_h=-\frac{i(\omega-\omega_c)}{f'(r_h)},\quad  a_0=\frac{\lambda+i\mu r_h}{r_h\sqrt{f'(r_h)}\,(2\rho_h+\tfrac12)},   
\end{aligned}
  \label{eq:horizonBC}
\end{equation}
with $\omega_c=-qA_t(r_h)$. The coefficient $a_0$ follows from inserting the ansatz into the first equation of \eqref{eq:radial} and using $iK/\Delta\simeq-\rho_h/(r-r_h)$. At infinity, the decaying solution is
\begin{equation}
  R_{1,2}\sim r^{\chi}e^{kr},\,\ k=-\sqrt{\mu^2-\omega^2},\,\
  \frac{R_2}{R_1}\to-\frac{i(k-i\omega)}{\mu},
  \label{eq:infBC}
\end{equation}
where the eigenvector follows from the constant-coefficient limit of \eqref{eq:radial}. Because the EEH corrections in \eqref{eq:asympt} start at $\mathcal{O}(r^{-5})$ in $A_t$ and $\mathcal{O}(r^{-6})$ in $f$, the exponent $\chi=[M(\mu^2-2\omega^2)+qQ\omega]/k$, the asymptotic potential
\begin{equation}
  V_{1,2}=\mu^2-\omega^2+\frac{2(qQ\omega-M\mu^2)}{r}+\mathcal{O}(r^{-2}),
\end{equation}
and therefore the far-field trapping condition
\begin{equation}
  M\mu^2-qQ\omega_R>0\quad\xrightarrow{\ \omega_R\simeq\mu\ }\quad\alpha_{\rm eff}\equiv M\mu-qQ>0
  \label{eq:trapping}
\end{equation}
are exactly those of RN~\cite{Huang2017,Zhang2026}. The leading hydrogenic spectrum $\omega_R\simeq\mu\left[1-\alpha_{\rm eff}^2/(2N^2)\right]$ with $N=n_r+\ell+1$ is also unchanged.

\subsection{A current identity: no real-frequency quasibound states}
\label{sec:current}

Classical superradiance is absent for Dirac fields~\cite{Unruh1976,IyerKumar1978}, and time-periodic normalizable Dirac solutions do not exist outside nonextremal black holes~\cite{Finster2000}. For the radial system \eqref{eq:radial} this can be shown in a few lines, for any $f$ and $A_t$ with a simple horizon. Differentiating $|R_{1,2}|^2$ and using \eqref{eq:radial},
\begin{align}
  \frac{d|R_1|^2}{dr}&=-\frac{2\operatorname{Im}K}{\Delta}|R_1|^2+\frac{2}{\sqrt\Delta}\operatorname{Re}\!\big[(\lambda+i\mu r)R_1^*R_2\big],\\
  \frac{d|R_2|^2}{dr}&=+\frac{2\operatorname{Im}K}{\Delta}|R_2|^2+\frac{2}{\sqrt\Delta}\operatorname{Re}\!\big[(\lambda-i\mu r)R_2^*R_1\big].
\end{align}
The two mixing terms are complex conjugates and cancel in the difference. Since $\operatorname{Im}K=r^2\omega_I$,
\begin{equation}
  \frac{d}{dr}\Big(|R_1|^2-|R_2|^2\Big)=-\frac{2\omega_I}{f(r)}\Big(|R_1|^2+|R_2|^2\Big).
  \label{eq:current}
\end{equation}
For real $\omega$ the right-hand side vanishes, so $|R_1|^2-|R_2|^2$ is constant. It vanishes at infinity for a bound state, while \eqref{eq:horizonBC} gives $|R_1|^2-|R_2|^2\to-|C_h|^2$ at the horizon, where $C_h$ is the amplitude of $R_2$. Hence $C_h=0$ and the solution is trivial. There are therefore no real-frequency quasibound states, for any $(a,Q,\mu,q,\lambda)$, including the threshold $\omega=\omega_c$, where $\rho_h=0$ and $R_2\to C_h$. A branch of quasibound frequencies followed continuously in parameter space can never cross the real axis, so $\omega_I<0$ on every branch connected to a damped mode. Equation \eqref{eq:current} also has a direct interpretation: $|R_2|^2$ at the horizon is the ingoing flux, which can only drain the cloud. We use the absence of sign changes of $\omega_I$ as a consistency check on all numerical branches.

\subsection{Far-zone estimate of the EEH shift}
\label{sec:farzone}

Because the EEH and RN potentials agree through $\mathcal{O}(r^{-2})$, the fine-structure correction that produced the leading ABG--RN splitting in Ref.~\cite{Zhang2026} is absent here. The first EEH term in the Schr\"odinger potential comes from $-K^2/r^4$ with $K=\omega r^2-qQr+qaQ^3/(10r^3)$:
\begin{equation}
  \delta V=-\frac{qaQ^3\omega}{5\,r^5}+\mathcal{O}(a\,r^{-6}).
  \label{eq:deltaV}
\end{equation}
For a hydrogenic state, the centrifugal term of the Dirac potential is $\lambda(\lambda+1)/r^2=\ell(\ell+1)/r^2$, so the bound-state wave function behaves as $\psi\propto r^{\ell+1}$ for $r\ll(\mu\alpha_{\rm eff})^{-1}$. The expectation value $\langle r^{-5}\rangle_{N\ell}$ converges at small $r$ only if $5<2\ell+3$, i.e.\ $\ell\ge2$. For the low-$\ell$ modes studied here, the integral is therefore dominated by the near zone $r=\mathcal{O}(M)$, and first-order far-zone perturbation theory does not apply. Using the small-$r$ normalization of the hydrogenic wave function, $|\psi_{N\ell}|^2\propto(\mu\alpha_{\rm eff})^{2\ell+3}r^{2\ell+2}/N^3$, together with $\delta\omega_R\simeq\langle\delta V\rangle/(2\mu)$, gives the scaling estimate
\begin{equation}
  M\big(\omega_R^{\rm EEH}-\omega_R^{\rm RN}\big)\;\propto\;\frac{a}{M^2}\,\frac{(M\mu\,\alpha_{\rm eff})^{2\ell+3}}{N^3},
  \label{eq:scaling}
\end{equation}
with an $\mathcal{O}(1)$ coefficient set by the near-horizon geometry. Equation \eqref{eq:scaling} predicts a linear real-frequency shift in $a$, suppressed by roughly two extra powers of $M\mu\alpha_{\rm eff}$ per unit $\ell$. Since the coefficient is itself a near-zone quantity that depends on $\omega$, $\lambda$ and the horizon data, \eqref{eq:scaling} should be read as an order-of-magnitude guide rather than a power law; Sec.~\ref{sec:results} shows where it holds and where it fails. By contrast, the damping rate is controlled by transmission through the inner barrier and need not follow \eqref{eq:scaling}.

%========================================================================
\section{Numerical method}
\label{sec:method}
%========================================================================

We solve \eqref{eq:radial} as a complex eigenvalue problem by two-sided shooting. For a trial $\omega$, we start the horizon solution at $r_0=r_h+\epsilon_h$ from the ingoing branch \eqref{eq:horizonBC} and integrate outward to a matching radius $r_m$. The decaying solution \eqref{eq:infBC} is started at $r_\infty=\max(80M,\,40/|\operatorname{Re}k|)$ and integrated inward. Integrating inward is numerically stable because the physical solution grows toward smaller $r$ while any admixture of the growing solution decays by $e^{-2|k|(r_\infty-r_m)}$. A quasibound frequency is a zero of
\begin{equation}
  \mathcal{D}(\omega)=\frac{R_1^{\rm hor}(r_m)}{R_2^{\rm hor}(r_m)}-\frac{R_1^{\infty}(r_m)}{R_2^{\infty}(r_m)} ,
  \label{eq:D}
\end{equation}
which, unlike the normalized Wronskian used in Ref.~\cite{Zhang2026}, is a holomorphic function of $\omega$. Holomorphy allows us to locate roots with Muller's method~\cite{Muller1956}, which converges in a few iterations even for $|\omega_I|\sim10^{-9}$. As an independent acceptance criterion we evaluate the scale-invariant Wronskian $W=(R_1^{\rm hor}R_2^\infty-R_2^{\rm hor}R_1^\infty)/(|R_1^{\rm hor}||R_2^\infty|+|R_2^{\rm hor}||R_1^\infty|)$ at the root and require $|W|<10^{-9}$, together with $\omega_I<0$ and $\omega_R<\mu$. We integrate the ODEs using an eighth-order Dormand--Prince scheme with a relative tolerance of $10^{-12}$. The horizon offset is $\epsilon_h=\min[10^{-6}M,\,10^{-3}(r_h-r_2)]$, where $r_2$ is the next root of $P(r)$; this keeps the starting point well inside the radius of convergence of the Frobenius series when the horizons approach each other near extremality.

\paragraph{Convergence.} For every entry of Table~\ref{tab:spectrum} we recomputed the root with $r_m\to r_m+10M$ and the ODE tolerance tightened to $10^{-13}$, and separately with $\epsilon_h\to10^{-7}M$. The frequencies change by at most $8\times10^{-13}$ and $6\times10^{-12}$ in $M\omega$, respectively, far below the digits quoted.

\paragraph{Benchmarks.} Setting $a=0$, the code reproduces the RN quasibound spectrum used as the baseline in Ref.~\cite{Zhang2026}. For $M\mu=0.4$, $qM=0.2$ and $Q/M=0.5$ we obtain $M\omega=0.3805065-0.0146337i$ for the fundamental $(n_r,j,\ell,\lambda)=(0,\tfrac12,0,-1)$ mode, in agreement with the published value $0.380507-0.014634i$. Replacing the background with ABG reproduces all five modes of Table~1 of Ref.~\cite{Zhang2026} to the quoted precision; for example, we find $0.3817935-0.0039401i$ for the ABG fundamental mode and $0.3955053-1.5775\times10^{-7}i$ for the $(0,\tfrac32,1,-2)$ mode.

\paragraph{Continuation.} Branches are followed in $a$, $Q$ or $\mu$ with fixed steps. Each new guess is obtained by linearly extrapolating $\omega_R/\mu$ and $\ln|\omega_I|$ from the two previous points. A branch is terminated when a root fails the acceptance tests, when the frequency jumps by more than $1\%$ of $\mu$, when $f'(r_h)<0.03$ (very close to extremality, where $\epsilon_h$ must become impractically small), or when $\varepsilon_h>0.5$.

%========================================================================
\section{Frequency-domain results}
\label{sec:results}
%========================================================================

Unless stated otherwise, we use $M=1$, $M\mu=0.4$ and $qM=0.2$, and we label modes by $(n_r,j,\ell,\lambda)$ and by the hydrogenic principal number $N=n_r+\ell+1$. We compare EEH and RN at identical $(Q,\mu,q)$ and labels through
\begin{equation}
  \Delta_R\equiv M\big(\omega_R^{\rm EEH}-\omega_R^{\rm RN}\big),\qquad
  \mathcal{R}_I\equiv\frac{|\omega_I^{\rm EEH}|}{|\omega_I^{\rm RN}|},
\end{equation}
so $\mathcal{R}_I<1$ means that the EEH mode lives longer than its RN counterpart.

\subsection{Background geometry}

\begin{figure*}[t]
  \centering
  \includegraphics[width=0.8\textwidth]{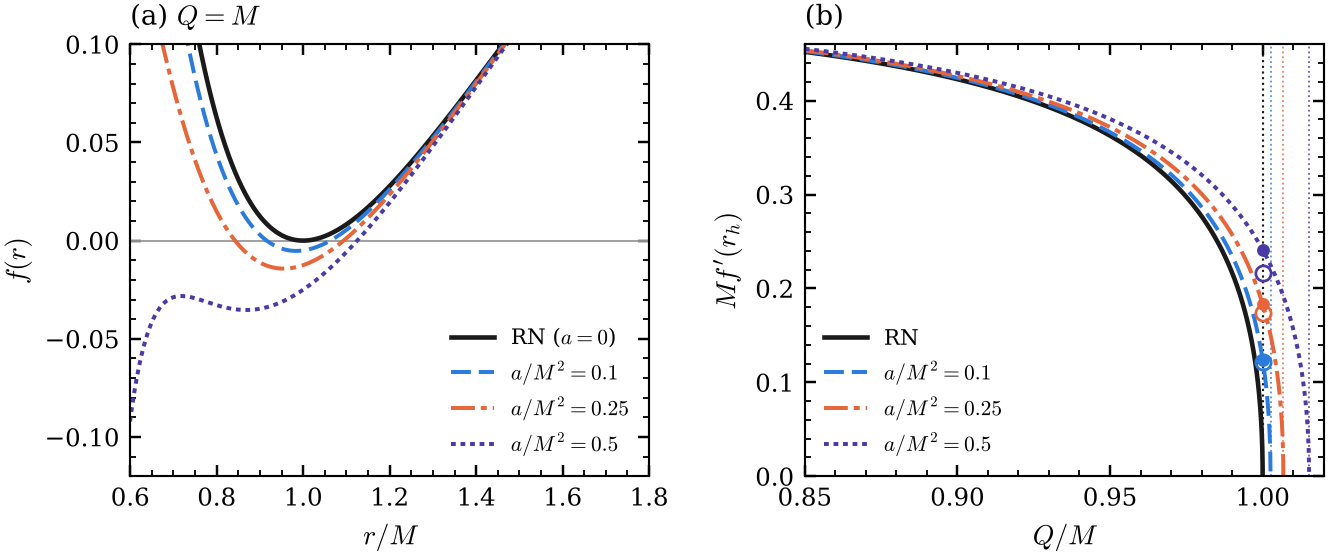}
  \caption{EEH horizon structure. (a) Metric function at $Q=M$ for RN and three values of the EH coupling. (b) $Mf'(r_h)$ as a function of $Q/M$. Open circles: the near-extremal expansion~\eqref{eq:QeqM} at $Q=M$. Vertical ticks: the extremal charges $Q_{\rm ext}$ from Eqs.~\eqref{eq:rext}--\eqref{eq:aext}.}
  \label{fig:background}
\end{figure*}

Figure~\ref{fig:background} summarizes the horizon structure of Sec.~\ref{sec:horizons}. At $Q=M$ [panel (a)], the RN function has a double root at $r=M$, while the EH term pushes $f$ below zero and opens a nondegenerate horizon at $r_h>M$. For $a/M^2=0.5$, the local maximum of $f$ inside the horizon is negative, so the inner horizon disappears and the causal structure becomes Schwarzschild-like. Panel (b) shows $Mf'(r_h)$. Away from extremality, the EEH and RN curves are indistinguishable. Near $Q=M$, the EEH surface gravity remains finite (open circles: the expansion \eqref{eq:QeqM}) and vanishes only at $Q_{\rm ext}>M$ (vertical ticks), in agreement with \eqref{eq:Qext}.

\subsection{Reference spectrum}

Table~\ref{tab:spectrum} lists the lowest modes at $Q/M=0.99$, for which $qQ=0.198$, $\alpha_{\rm eff}=0.202$, and the EEH horizon is $r_h/M=1.1797$ for $a/M^2=0.5$ (RN: $1.1411$). Three features stand out.
\begin{itemize}
\item The real frequencies agree with RN to within $6.1\times10^{-5}$, and the shift falls by three orders of magnitude from the $\ell=0$ to the $\ell=1$ modes.
\item The damping ratio depends on the branch. For $\ell=0$ the EEH modes live slightly \emph{longer} ($\mathcal{R}_I\simeq0.98$ at $a/M^2=0.5$), while for $\ell=1$ they decay \emph{faster}.
\item The $(0,\tfrac32,1,-2)$ mode is by far the most sensitive: its damping rate grows by $40\%$ at $a/M^2=0.5$.
\end{itemize}
The hydrogenic degeneracy of the three $N=2$ states is lifted already in RN; the EH correction does not change the ordering.

\begin{table*}[t]
\centering
\caption{Quasibound spectrum at $Q/M=0.99$, $M\mu=0.4$, $qM=0.2$. The EEH column uses $a/M^2=0.5$; $\mathcal{R}_I$ is given for three values of the EH coupling. All digits shown are stable under the convergence tests of Sec.~\ref{sec:method}.}
\label{tab:spectrum}
\setlength{\tabcolsep}{3.5pt}
\resizebox{\textwidth}{!}{%
\begin{tabular}{ccccr ll r ccc}
\hline\\
\toprule
$N$ & $n_r$ & $j$ & $\ell$ & $\lambda$ & $M\omega^{\rm RN}$ & $M\omega^{\rm EEH}$ & $\Delta_R$ & \multicolumn{3}{c}{$\mathcal{R}_I$} \\
%\cmidrule(lr){9-11}
\hline\\
 & & & & & & ($a/M^2=0.5$) & & $0.1$ & $0.25$ & $0.5$ \\
\midrule
1 & 0 & $\tfrac12$ & 0 & $-1$ & $0.38865639-2.18298\times10^{-3}i$ & $0.38871748-2.14710\times10^{-3}i$ & $6.11\times10^{-5}$ & 0.9962 & 0.9909 & 0.9836\\[2pt]
2 & 1 & $\tfrac12$ & 0 & $-1$ & $0.39733425-3.59836\times10^{-4}i$ & $0.39734163-3.53416\times10^{-4}i$ & $7.39\times10^{-6}$ & 0.9960 & 0.9905 & 0.9822\\
2 & 0 & $\tfrac12$ & 1 & $+1$ & $0.39772862-3.23180\times10^{-6}i$ & $0.39772871-3.28238\times10^{-6}i$ & $8.53\times10^{-8}$ & 1.0027 & 1.0072 & 1.0157\\
2 & 0 & $\tfrac32$ & 1 & $-2$ & $0.39787661-1.40387\times10^{-9}i$ & $0.39787660-1.96478\times10^{-9}i$ & $-3.69\times10^{-9}$ & 1.0805 & 1.2006 & 1.3996\\[2pt]
3 & 1 & $\tfrac12$ & 1 & $+1$ & $0.39900765-1.21403\times10^{-6}i$ & $0.39900769-1.23169\times10^{-6}i$ & $3.18\times10^{-8}$ & 1.0025 & 1.0067 & 1.0145\\
\bottomrule
\hline
\end{tabular}}
\end{table*}

\subsection{Dependence on the Euler--Heisenberg coupling}

\begin{figure*}[t]
  \centering
  \includegraphics[width=0.8\textwidth]{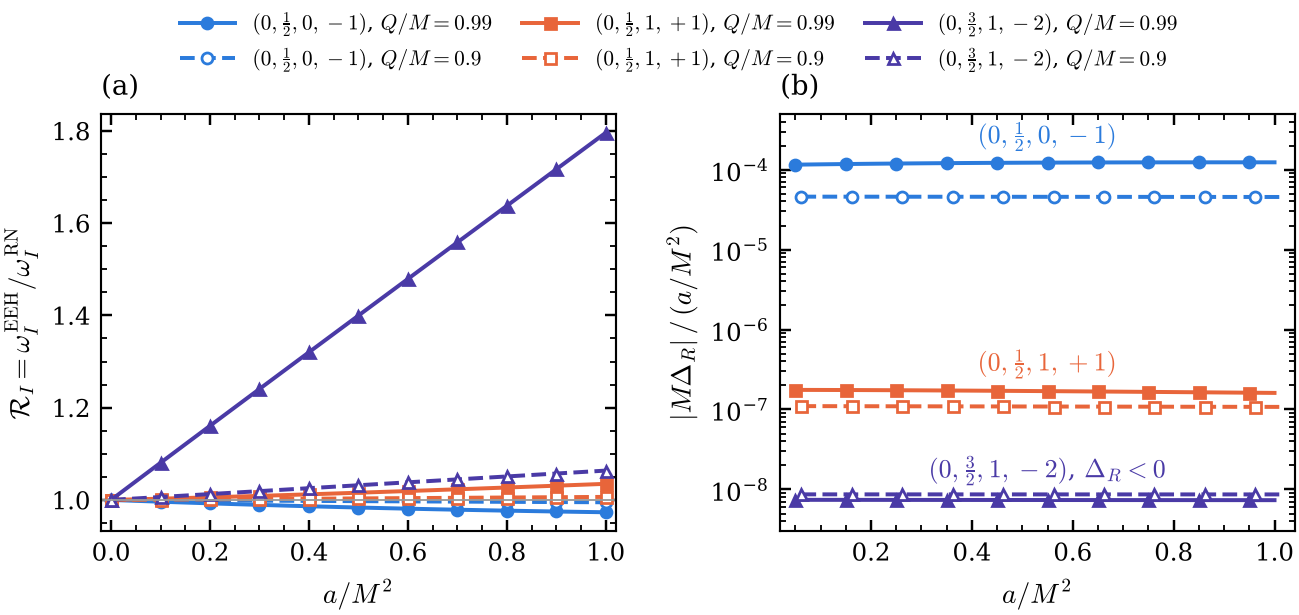}
  \caption{Dependence on the EH coupling for the three fundamental branches at $M\mu=0.4$, $qM=0.2$. Filled markers and solid lines: $Q/M=0.99$; open markers and dashed lines: $Q/M=0.9$. (a) Damping ratio $\mathcal{R}_I$. (b) Absolute value of the real-frequency shift divided by $a$; the $(0,\tfrac32,1,-2)$ shifts (lowest set) are negative.}
  \label{fig:ascan}
\end{figure*}

Figure~\ref{fig:ascan} follows the three $N\le2$ fundamental branches with $a/M^2\in[0,1]$ at $Q/M=0.99$ and $0.9$. Both $\mathcal{R}_I-1$ and $\Delta_R$ are linear in $a$ to good accuracy over the whole range [panel (b) shows that $\Delta_R/a$ is constant]. This confirms that, at fixed $Q<M$, the spectrum responds to the EH coupling at first order. At $a/M^2=1$ and $Q/M=0.99$ the $(0,\tfrac32,1,-2)$ damping rate is enhanced by $80\%$ ($\mathcal{R}_I=1.796$), compared with $+3.5\%$ for $(0,\tfrac12,1,+1)$ and $-2.7\%$ for the ground state. Moving away from extremality to $Q/M=0.9$ reduces all effects by a factor of $5$--$12$, as expected from the $Q^4$ and $Q^3$ scaling of the EH terms in \eqref{eq:fEEH}--\eqref{eq:AtEEH} and from the larger horizon.

The real-frequency shifts in Fig.~\ref{fig:ascan}(b) are ordered as predicted by \eqref{eq:scaling}. The $\ell=0$ shift $\Delta_R/(a/M^2)\simeq1.2\times10^{-4}$ exceeds the $\ell=1$ shifts by three to four orders of magnitude; the factor $(M\mu\alpha_{\rm eff})^2/N^3$ alone accounts for about $10^{-3}$. The $(0,\tfrac32,1,-2)$ shift is negative, showing that the $\mathcal{O}(1)$ near-zone coefficient in \eqref{eq:scaling} depends on the spinor branch and not only on $\ell$.

\subsection{What controls the damping: metric versus potential}

The EH term enters the radial system \eqref{eq:radial} in two places: through $f$ in $\Delta$ and the tortoise coordinate, and through $A_t$ in $K$. To separate them, we solve \eqref{eq:radial} with only one correction switched on. This is a diagnostic background, not a solution of the field equations. Table~\ref{tab:decomposition} shows the result at $Q/M=0.99$, $a/M^2=0.5$. The two contributions add up to within $0.4\%$ of the full shift, consistent with the linear response seen in Fig.~\ref{fig:ascan}.

\begin{itemize}
\item \emph{Potential correction.} The term $+aQ^3/(10r^5)$ reduces $|A_t|$ near the horizon, i.e.\ it screens the Coulomb repulsion between the hole and the like-charged field ($qQ>0$). For all three branches it \emph{increases} the damping by $1$--$6\%$: with less repulsion, more of the cloud reaches the horizon.
\item \emph{Metric correction.} The term $-aQ^4/(20r^6)$ moves the horizon outward and raises $f'(r_h)$. Its effect has opposite signs for $\ell=0$ and $\ell=1$. It \emph{reduces} the ground-state damping by $2.8\%$, which dominates the net $-1.6\%$ shift, and it dominates the $(0,\tfrac32,1,-2)$ enhancement ($+33\%$ of the total $+40\%$).
\end{itemize}

\begin{table}[t]
\centering
\caption{Separation of the EH effect at $Q/M=0.99$, $M\mu=0.4$, $qM=0.2$, $a/M^2=0.5$. ``Metric'' keeps the EEH $f(r)$ with the Coulomb potential; ``potential'' keeps the RN metric with the EEH $A_t(r)$. The last three columns give the WKB integral \eqref{eq:wkb} of the inner barrier for RN and EEH and the resulting estimate $e^{-2\Delta S}$ of $\mathcal{R}_I$.}
\label{tab:decomposition}
\small
\begin{tabular}{l ccc ccc}
\toprule
 & \multicolumn{3}{c}{$\mathcal{R}_I$ (exact)} & \multicolumn{3}{c}{WKB} \\
%\cmidrule(lr){2-4}\cmidrule(lr){5-7}
mode & metric & potential & full & $S_{\rm RN}$ & $S_{\rm EEH}$ & $e^{-2\Delta S}$ \\
\midrule
$(0,\tfrac12,0,-1)$ & 0.9718 & 1.0107 & 0.9836 & 0.242 & 0.240 & 1.003 \\
$(0,\tfrac12,1,+1)$ & 0.9983 & 1.0174 & 1.0157 & 2.314 & 2.308 & 1.010 \\
$(0,\tfrac32,1,-2)$ & 1.3329 & 1.0629 & 1.3996 & 5.964 & 5.829 & 1.312 \\
\bottomrule
\end{tabular}
\end{table}

\begin{figure*}[t]
  \centering
  \includegraphics[width=0.8\textwidth]{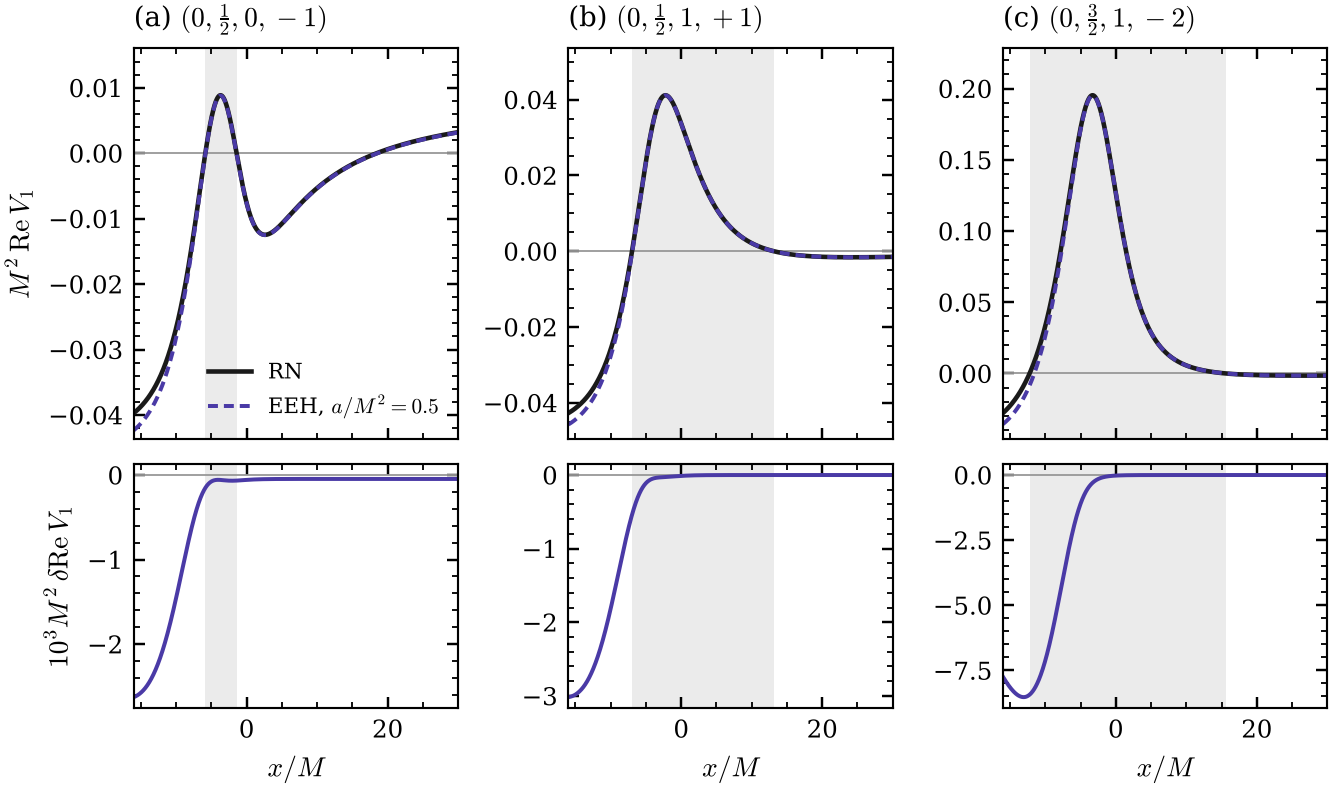}
  \caption{Top: real part of the effective potential $V_1$, Eq.~\eqref{eq:V1}, at the hydrogenic frequency for the three fundamental branches at $Q/M=0.99$, $M\mu=0.4$, $qM=0.2$. Solid: RN; dashed: EEH with $a/M^2=0.5$. Bottom: the difference $\delta\operatorname{Re}V_1=\operatorname{Re}V_1^{\rm EEH}-\operatorname{Re}V_1^{\rm RN}$ at equal $r-r_h$.}
  \label{fig:potential}
\end{figure*}

A simple barrier picture accounts for the $\ell=1$ branches. Figure~\ref{fig:potential} compares $\operatorname{Re}V_1$, evaluated at the hydrogenic frequency, for RN and EEH. The EH term raises the potential near the horizon but lowers and narrows the barrier just outside it. We quantify the change with the WKB tunneling integral of the innermost barrier,
\begin{equation}
  S=\int_{\operatorname{Re}V_1>0}\sqrt{\operatorname{Re}V_1}\;dx ,
  \label{eq:wkb}
\end{equation}
and the transmission estimate $\mathcal{R}_I\simeq e^{-2(S_{\rm EEH}-S_{\rm RN})}$. For the $(0,\tfrac32,1,-2)$ mode the barrier is opaque ($S\simeq6$), and the estimate $1.31$ reproduces most of the exact enhancement $1.40$. For $(0,\tfrac12,1,+1)$ ($S\simeq2.3$) it gives $1.010$ against $1.016$. The estimate fails for the ground state: its barrier is almost transparent ($S\simeq0.24$), so the damping is controlled by the amplitude of the cloud near the horizon rather than by tunneling, and the sign of $\mathcal{R}_I-1$ follows the metric contribution instead. The strong $\ell$ dependence in Table~\ref{tab:spectrum} thus follows from the exponential sensitivity of tunneling through tall barriers: a fixed fractional change in the barrier produces a change in $\omega_I$ that grows with $S$.

\subsection{Charge dependence and the near-extremal regime}

\begin{figure*}[t]
  \centering
  \includegraphics[width=0.8\textwidth]{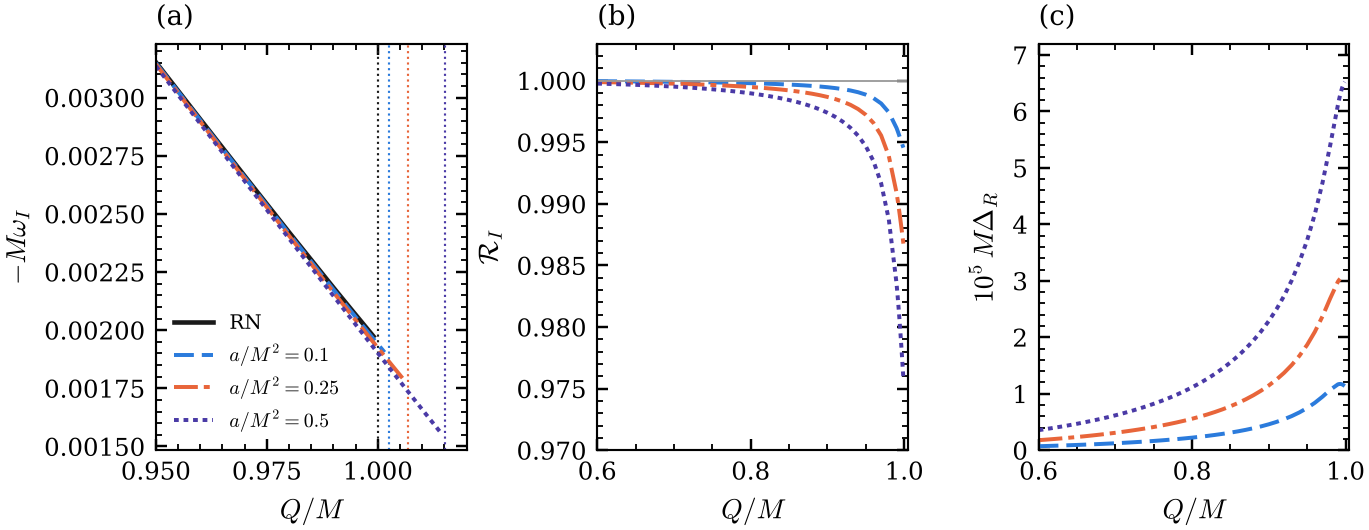}
  \caption{Ground state $(0,\tfrac12,0,-1)$ as a function of the charge, for $M\mu=0.4$, $qM=0.2$. (a) Damping rate near extremality; dotted vertical lines mark $Q=M$ (black) and the EEH extremal charges $Q_{\rm ext}(a)$ (colored). (b) Damping ratio and (c) real-frequency shift relative to RN at equal $Q$.}
  \label{fig:qscan}
\end{figure*}

Figure~\ref{fig:qscan} follows the ground state as a function of $Q/M$ for $a/M^2=0.1$, $0.25$ and $0.5$, from $Q/M=0.3$ up to the point where $Mf'(r_h)<0.03$. The RN branch ends at $Q/M=0.999$, where $Mf'(r_h)=0.08$. The EEH branches continue smoothly through $Q=M$ to within $Mf'(r_h)\approx0.04$--$0.08$ of the EEH extremal limit; for $a/M^2=0.5$ the last point is $Q/M=1.014$, beyond the RN bound. Throughout the range, $\omega_I$ stays negative and varies smoothly, consistent with Sec.~\ref{sec:current}. The ground-state damping is remarkably insensitive to the approach to extremality, even though $f'(r_h)$ drops by a factor of about three over the last $1\%$ in $Q$. This again reflects the transparent inner barrier of the $\ell=0$ state.

Panels (b) and (c) show that the EEH--RN differences are negligible for $Q/M\lesssim0.7$ and grow steeply as $Q\to M$. Just below $Q=M$, the deviation of $\mathcal{R}_I$ from unity grows steeply where the horizon data become nonanalytic in $a$ (Eq.~\eqref{eq:QeqM}). Even there, the ground state changes by only $\lesssim2.5\%$ in $\omega_I$ and $\lesssim7\times10^{-5}$ in $M\omega_R$ for $a/M^2\le0.5$.

\subsection{Dependence on the field mass}

\begin{figure*}[t]
  \centering
  \includegraphics[width=0.9\textwidth]{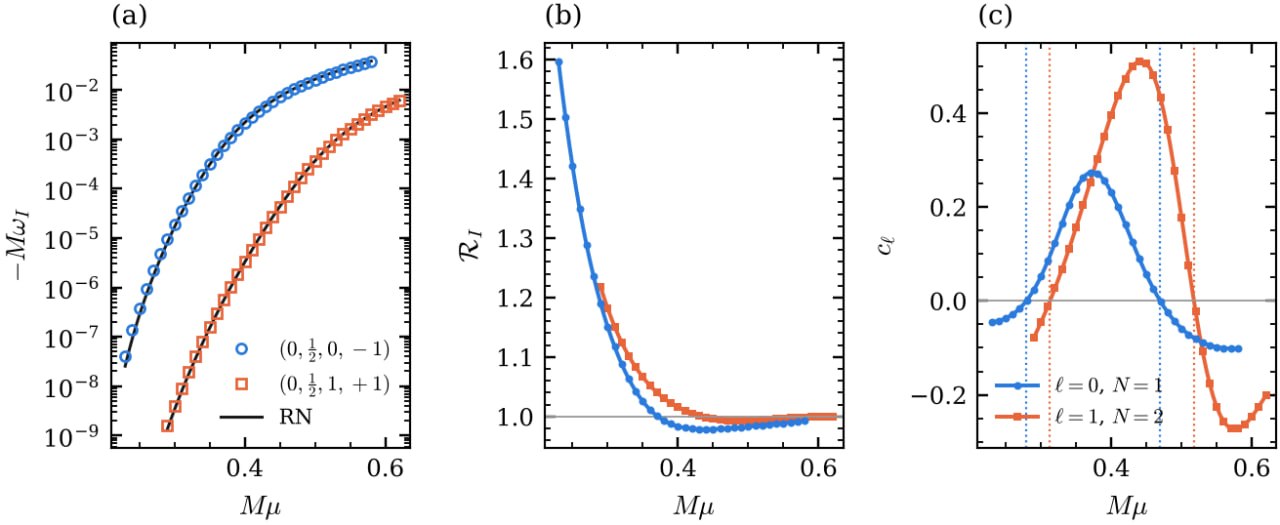}
  \caption{Dependence on the field mass at $Q/M=0.99$, $qM=0.2$. (a) Damping rates for RN (solid lines) and EEH with $a/M^2=0.5$ (markers). (b) Damping ratio and (c) real-frequency shift.}
  \label{fig:muscan}
\end{figure*}

Figure~\ref{fig:muscan} shows the $(0,\tfrac12,0,-1)$ and $(0,\tfrac12,1,+1)$ branches as functions of $M\mu$ at $Q/M=0.99$, for RN and for EEH with $a/M^2=0.5$. The scan covers $0.23\le M\mu\le0.58$ for $\ell=0$ and $0.29\le M\mu\le0.62$ for $\ell=1$; the lower ends approach the trapping threshold $M\mu=qQ=0.198$ of \eqref{eq:trapping}. The damping rates span more than six orders of magnitude, from $|M\omega_I|\simeq2.5\times10^{-8}$ near threshold to $3.8\times10^{-2}$ for the most compact $\ell=0$ cloud.

The damping ratio [panel (b)] is not monotonic, and changes sign relative to unity.
\begin{itemize}
\item \emph{Weak binding.} Near threshold the EEH modes decay considerably \emph{faster}: $\mathcal{R}_I=1.60$ at $M\mu=0.23$ for $\ell=0$ and $1.22$ at $M\mu=0.29$ for $\ell=1$.
\item \emph{Intermediate masses.} $\mathcal{R}_I$ crosses unity at $M\mu\simeq0.37$ ($\ell=0$) and $0.44$ ($\ell=1$) and reaches a shallow minimum of $0.979$ at $M\mu=0.44$ ($\ell=0$).
\item \emph{Large masses.} $\mathcal{R}_I$ returns toward unity as the cloud becomes compact.
\end{itemize}
The weak-binding enhancement has a simple origin. As $\alpha_{\rm eff}\to0$, the long-range Newtonian attraction and Coulomb repulsion nearly cancel, so the residual short-range structure controls the small fraction of the cloud that reaches the horizon, where the EH terms act. A fixed absolute change of the near-horizon potential then produces a large relative change of $\omega_I$. The trend mirrors the one found for ABG in Ref.~\cite{Zhang2026}, where the largest relative effects also occurred on the weak-binding side.

The real-frequency shift [panel (c)] changes sign twice for each branch: at $M\mu\simeq0.28$ and $0.47$ for $\ell=0$, and at $0.31$ and $0.52$ for $\ell=1$. At large $M\mu$ it is negative and grows to $|\Delta_R|=5.5\times10^{-4}$ at $M\mu=0.58$ for $\ell=0$. Between the zeros, the positive shift follows a power law. For $\ell=0$ a fit over $0.30\le M\mu\le0.45$ gives $\Delta_R\propto(M\mu\,\alpha_{\rm eff})^{3.2}$, consistent with the exponent $2\ell+3=3$ of \eqref{eq:scaling}. For $\ell=1$, the fitted exponent is $7.4$, steeper than the naive value $5$. This confirms that the scaling estimate captures the order of magnitude and the $\ell$ hierarchy, but not the detailed near-zone coefficient, which depends on $\omega$ through $K(r)$ and changes sign when the binding becomes strong.

%========================================================================
\section{Time-domain evolution}
\label{sec:timedomain}
%========================================================================

As an independent check, we evolve the radial Dirac system in the tortoise coordinate $x$,
\begin{equation}
\begin{aligned}
  (\partial_t+\partial_x)R_1=iqA_tR_1+\sqrt f\Big(i\mu+\frac{\lambda}{r}\Big)R_2,\\
  (\partial_t-\partial_x)R_2=iqA_tR_2+\sqrt f\Big(i\mu-\frac{\lambda}{r}\Big)R_1,
\end{aligned}
  \label{eq:td}
\end{equation}
which follows from \eqref{eq:radial} with $\omega\to i\partial_t$ and $d/dr=f^{-1}d/dx$. For a mode $e^{-i\omega t}$, Eq.~\eqref{eq:td} reproduces \eqref{eq:radial} exactly, so the time dependence of the numerical solution is directly $e^{-i\omega t}$ in the gauge $A_t(\infty)=0$, with no additional phase shift. We integrate \eqref{eq:td} along characteristics with a second-order Crank--Nicolson (trapezoidal) scheme on a grid with $\Delta t=\Delta x=h$, following Refs.~\cite{GundlachPricePullin1994,Zhang2026}. Each new grid value requires solving a $2\times2$ linear system. Evolving only inside the observer's causal diamond, $x\in[x_{\rm obs}-T,x_{\rm obs}+T]$, removes the need for artificial boundary conditions. The function $r(x)$ is obtained from $dr/dx=f$ with $r(10M)=10M$. For $x<-40M$, where $r-r_h<10^{-8}M$, we use the near-horizon form $r-r_h\propto e^{f'(r_h)x}$.

\begin{figure*}[t]
  \centering
  \includegraphics[width=0.8\textwidth]{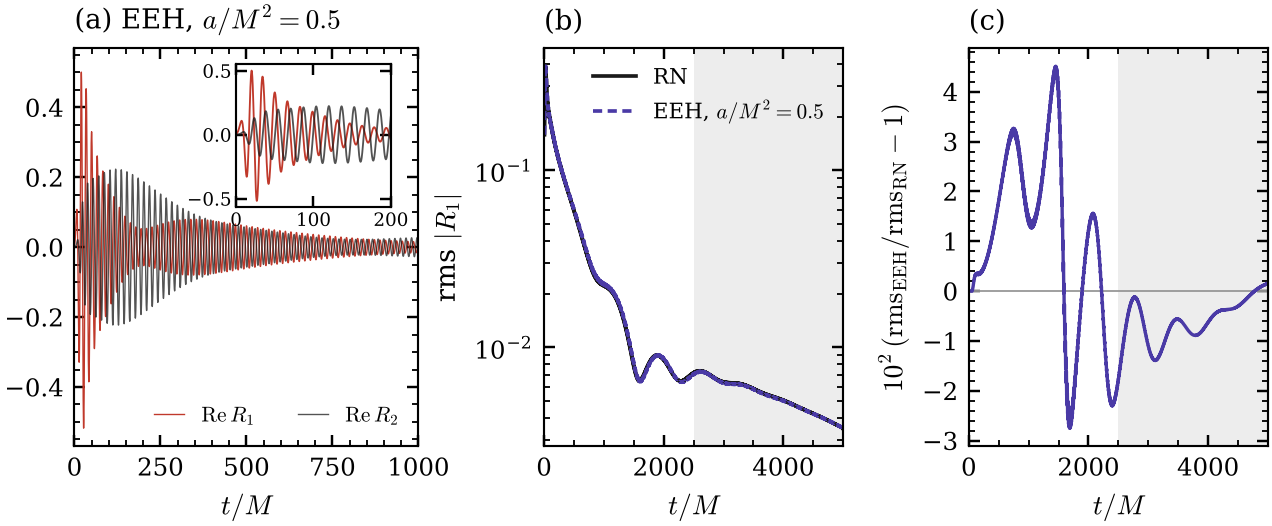}
  \caption{Time-domain evolution for $Q/M=0.99$, $M\mu=0.4$, $qM=0.2$, $\lambda=-1$, observed at $x_{\rm obs}=20M$. (a) EEH waveform for $a/M^2=0.5$; the inset shows the early response. (b) Running root-mean-square of $|R_1|$ over one period $2\pi/\mu$, for RN and EEH.}
  \label{fig:td}
\end{figure*}

The initial data are a Gaussian packet $R_1=\exp[-(x-x_g)^2/2\sigma^2]$, $R_2=0$, with $x_g=7M$ and $\sigma=4M$. We record the signal at $x_{\rm obs}=20M$, close to the peak of the ground-state cloud, so the $N=1$ mode dominates the quasibound stage. We use the same parameters as in Table~\ref{tab:spectrum} with $\lambda=-1$ and evolve to $T=5000M$. Figure~\ref{fig:td}(a) shows the EEH waveform: a prompt response is followed by a quasibound ringing in which $R_1$ and $R_2$ oscillate with the same frequencies but different amplitudes and phases. The envelope in panel (b) shows the decay of the ground state, the beating with the $N\ge2$ modes and, at late times, the slowly decaying massive tail. On the scale of the figure, the RN and EEH envelopes are indistinguishable, consistent with the percent-level differences in the frequency domain.

\paragraph{Mode extraction and discretization error.}
We extract complex frequencies from the late-time signal with the matrix-pencil method~\cite{HuaSarkar1990,SarkarPereira1995,Berti2007}. Before fitting, we multiply the signal by $e^{i\omega_0t}$ with $M\omega_0=0.392$ and resample every $8M$, moving the relevant band to low frequencies and avoiding aliasing. Two systematic effects must be controlled.

\begin{enumerate}
\item \emph{Discretization error.} The Crank--Nicolson scheme has a second-order phase error. With $h=0.5M$ the extracted real frequency is too low by $\simeq1.4\times10^{-3}/M$, and with $h=0.25M$ by about a quarter of that. We therefore run every case with $h=0.5M$ and $0.25M$ and use the Richardson estimate $\omega_R(0)\simeq\omega_R(h/2)+[\omega_R(h/2)-\omega_R(h)]/3$; the damping rates are much less sensitive and are taken from the $h=0.25M$ run. We validated this procedure on the ABG benchmark of Ref.~\cite{Zhang2026} ($Q/M=0.5$, $\lambda=+1$, $n_r=0$): the extrapolated value $M\omega_R=0.394996$ agrees with the frequency-domain value $0.395000$. For the evolution equations \eqref{eq:td}, the numerical solution already oscillates as $e^{-i\omega t}$ in the gauge $A_t(\infty)=0$, so no gauge correction is applied. Any residual offset at finite $h$ is a discretization effect that vanishes as $h^2$.
\item \emph{Fitting systematics.} The fitted frequencies depend on the fitting window and the number of modes, because the signal also contains the $N\ge2$ modes and the power-law tail. We repeat the fit for seven windows between $t=150M$ and $3500M$, with three to five modes, and quote the mean and standard deviation.
\end{enumerate}

Table~\ref{tab:td} compares the results with the frequency domain. The absolute frequencies agree to $1\times10^{-4}$ in $M\omega_R$ and about $1\%$ in $\omega_I$, well within the spread between fitting windows ($2\times10^{-4}$ and $6\%$). The EEH--RN \emph{differences} are much more robust: the window-dependent systematics are almost identical for the two backgrounds and cancel. The time-domain estimates $\Delta_R=(6.1\pm1.0)\times10^{-5}$ and $\mathcal{R}_I=0.982\pm0.004$ agree with the frequency-domain values $6.11\times10^{-5}$ and $0.9836$. The time-domain evolution thus confirms both the mode content and the EH-induced shifts.

\begin{table*}[t]
\centering
\caption{Frequency-domain (FD) versus time-domain (TD) ground state $(0,\tfrac12,0,-1)$ at $Q/M=0.99$, $M\mu=0.4$, $qM=0.2$. TD values are Richardson-extrapolated in $h$ and averaged over seven matrix-pencil fitting windows (mean $\pm$ standard deviation).}
\label{tab:td}
\small
\begin{tabular}{lll}
\toprule
\hline\\
 & FD & TD \\
 \hline\\
\midrule
$M\omega^{\rm RN}$ & $0.388656-2.183\times10^{-3}i$ & $(0.38876\pm0.00020)-(2.21\pm0.14)\times10^{-3}i$ \\
$M\omega^{\rm EEH}$ ($a/M^2=0.5$) & $0.388717-2.147\times10^{-3}i$ & $(0.38882\pm0.00020)-(2.17\pm0.14)\times10^{-3}i$ \\
$\Delta_R$ & $6.11\times10^{-5}$ & $(6.1\pm1.0)\times10^{-5}$ \\
$\mathcal{R}_I$ & $0.9836$ & $0.982\pm0.004$ \\
\bottomrule
\hline
\end{tabular}
\end{table*}

%========================================================================
\section{Summary and discussion}
\label{sec:conclusions}
%========================================================================

We have studied quasibound states of a massive charged Dirac field on the Einstein--Euler--Heisenberg black hole, the minimal extension of Reissner--Nordstr\"om by the leading QED vacuum-polarization term. The main results are as follows.

\begin{enumerate}
\item \emph{Background.} To first order in the EH coupling the solution is $f=1-2M/r+Q^2/r^2-aQ^4/(20r^6)$ and $A_t=-Q/r+aQ^3/(10r^5)$. Vacuum polarization raises the extremal charge to $Q_{\rm ext}=M[1+a/(40M^2)+23a^2/(3200M^4)+\dots]$ and gives the $Q=M$ configuration a nondegenerate horizon, with surface-gravity parameter $Mf'(r_h)\simeq\sqrt{a/5M^2}$.

\item \emph{No real-frequency modes.} The identity \eqref{eq:current} shows that a static background of this class cannot support real-frequency quasibound Dirac states. Damped branches therefore never cross into instability, and all modes found are damped.

\item \emph{Far zone.} Because the EH corrections enter the far zone only at $\mathcal{O}(r^{-5})$, the trapping condition and the hydrogenic spectrum are exactly those of RN. Unlike the regular ABG case~\cite{Zhang2026}, fine-structure order does not split the levels. The real-frequency shifts are near-zone effects: they are linear in $a$, of order $10^{-5}$--$10^{-4}$ for $\ell=0$ at $M\mu=0.4$ (reaching $5\times10^{-4}$ for the most compact clouds), and suppressed by roughly three orders of magnitude for $\ell=1$.

\item \emph{Damping.} The damping rates respond more strongly and in a branch-dependent way. At $Q/M=0.99$ the $\ell=0$ ground state lives slightly longer than in RN (by $1.6\%$ at $a/M^2=0.5$), while the $(0,\tfrac32,1,-2)$ mode decays $40\%$ faster at $a/M^2=0.5$ and $80\%$ faster at $a/M^2=1$. Separating the metric and potential corrections shows two mechanisms. Screening the Coulomb repulsion always increases leakage. The metric modification reduces it for $\ell=0$ and strongly increases it for $\ell=1$ and $j=3/2$ by lowering and narrowing the inner barrier, as a WKB estimate confirms. Near the trapping threshold, the relative effect grows further, up to $\mathcal{R}_I=1.6$.

\item \emph{Time domain.} Time-domain evolutions confirm the frequency-domain spectrum. Differential EEH--RN quantities are reproduced within the fitting-window spread ($\sim15\%$ in $\Delta_R$ and $0.4\%$ in $\mathcal{R}_I$), once the second-order Crank--Nicolson phase error is removed by Richardson extrapolation.
\end{enumerate}

Physically, these results make concrete the expectation that vacuum polarization leaves almost no imprint on the oscillation frequencies of a fermionic cloud but can change its lifetime. The change is largest for higher angular momentum and for weakly bound clouds, which are also the longest-lived. In terms of the conversion of Ref.~\cite{Zhang2026}, the lifetime of the $(0,\tfrac32,1,-2)$ state at our reference point, $\tau=1/|\omega_I|$, changes from $7.1\times10^8M$ to $5.1\times10^8M$ at $a/M^2=0.5$.

As we stressed in Sec.~\ref{sec:validity}, the parameter range needed for visible effects corresponds to near-extremal charges, far above the Schwinger threshold for astrophysical masses. The results should therefore be regarded as a controlled probe of how a weak-field NLED correction to the exterior geometry shapes fermionic spectra. The same method applies directly to other NLED models with a Maxwell weak-field limit, such as Born--Infeld~\cite{BornInfeld1934,MaLi2020}. Natural extensions include higher orders in $a$, which require solving the cubic constraint \eqref{eq:gauss} exactly; the backreaction of the Dirac cloud; and rotating EEH black holes~\cite{BretonLammerzahlMacias2019}, where frame dragging and the EH correction compete.

\bibliographystyle{apsrev4-1}
\bibliography{references}
\end{document}